\documentclass[twocolumn,pra]{revtex4}% Physical Review A
\usepackage{graphicx}% Include figure files
\usepackage{dcolumn}% Align table columns on decimal point
\usepackage{bm}% bold math
\begin{document}

\preprint{APS/123-QED}

\title{Probing dipolar effects with condensate shape oscillation}
\author{S. Yi and L. You}
\affiliation{School of Physics, Georgia Institute of Technology,
Atlanta, GA 30332-0430}

\date{\today}

\begin{abstract}
We discuss the low energy shape oscillations of a magnetic trapped
atomic condensate including the spin dipole interaction.
When the nominal isotropic s-wave interaction strength becomes
tunable through a Feshbach resonance (e.g. as for $^{85}$Rb atoms),
anisotropic dipolar effects are shown to be detectable
under current experimental conditions
[E. A. Donley {\it et al.}, Nature {\bf 412}, 295 (2001)].
\end{abstract}

\maketitle
%\narrowtext

Collective excitations play an important role in probing
microscopic interactions \cite{ph}.
The recently available gaseous atomic Bose-Einstein
condensates (BEC), have proven to be a profitable
testing ground for such studies \cite{castin}.
Atomic BEC are dilute with properties completely determined
by binary interactions.
Experimentally they are created at very low temperatures
when the short range atom-atom (collision) interaction can be
described by a single parameter: the s-wave scattering
length $a_{\rm sc}$. All higher order partial wave
collisions are suppressed at zero collision energy in
a short ranged potential. This enable atoms to be
modelled as hard spheres of radius $a_{\rm sc}$,
reflecting the isotropic ground state interaction.
Inside an electric or magnetic field, however,
ground state atoms may be polarized, e.g.
in a magnetic trap the direction of atomic spins
(of the valance electron for alkali) becomes aligned.
The resulting dipole interaction between condensed
atom pairs many not simply average out.
In this article, we investigate such spin dipole
effects on shape oscillation frequencies of an atomic condensate.
Although dipolar effects are typically small compared to
the dominant s-wave contact interaction, our study
indicates that these shifts
become observable in the $^{85}$Rb BEC \cite{e2,e3},
when a Feshbach resonance is used to tune
$a_{\rm sc}$ near zero \cite{e2}.

\begin{figure}
\includegraphics[width=1.5in]{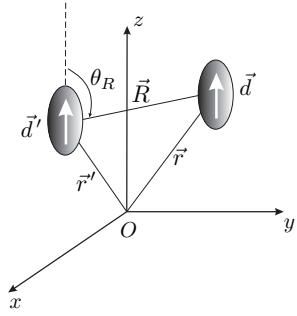}
\caption{Geometry for the interaction of two aligned dipoles.}
\label{fig1}
\end{figure}

Dipolar interaction in atomic BEC leads to physics beyond the
usual s-wave contact term, mainly because of the
modified low energy collision threshold behavior due to
the anisotropic nature of the interaction \cite{mircea}.
For magnetic spin dipoles, the net effect in the dilute
gas sample is rather small, one can therefor approximate the
complete two-body interaction by \cite{yi1,goral}
\begin{eqnarray}
V(\vec R)=g_0\delta(\vec R)+g_2{1-3\cos^2\theta_R\over R^3}
\end{eqnarray}
where $\vec R=\vec r-\vec r\,'$, and $g_0=4\pi\hbar^2a_{\rm sc}/M$
is the s-wave contact term. More generally, the dipole strength
$g_2$ equals $\alpha^2(0){\cal E}^2$ ($\alpha(0)$ atomic polarizability)
or $\mu^2$ ($\mu$ magnetic dipole moment) respectively for electric
or magnetic dipoles.
Figure \ref{fig1} illustrates the geometry for two
aligned dipoles along the local magnetic
field (z-axis). Several interesting properties of a dipolar
condensate have already been discussed \cite{yi1,goral,santos,yi2}.
In this paper, we focus on the shape oscillations
of a trapped dipolar condensate assuming a tunable
$a_{\rm sc}$.

\begin{figure}
\includegraphics[width=2.25in]{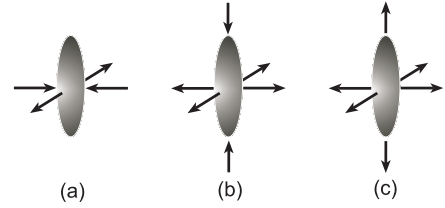}
\caption{Collective excitations of a
condensate with cylindrical symmetry.
Mode a is purely radial due to the cylindrical symmetry.
It's angular momentum projection along the z-axis $m=2$. Mode b
and c are respectively the quadrupole and monopole oscillations.
They are customarily called the low and high $m=0$ modes.}
\label{fig2}
\end{figure}

In the standard approach, low energy collective excitations
are described by the Bogoliubov theory \cite{stringari}.
The condensate wavefunction $\phi(\vec r,t)$ is governed
by the Gross-Pitaevskii equation, while the non-condensed atoms
are described by quasi-particles \cite{keith}, some of
which have been studied experimentally \cite{jila,mit}.
The inclusion of the non-local dipolar interaction makes
the Bogoliubov approach impractical to implement numerically.
We therefore will rely on two alternative methods
to study the three characteristic shape modes \cite{jila}
as graphically illustrated for a cylindrical
symmetric trap in Fig. \ref{fig2}.

First we adopt the highly successful time-dependent variation
approach used in Ref. \cite{zoller} by assuming a Gaussian ansatz
\begin{eqnarray}
\phi(x,y,z,t)=A(t)\prod_{\eta=x,y,z}e^{-\eta^2/2q^2_\eta+i\eta^2\beta_\eta(t)}.
\label{gausan}
\end{eqnarray}
For an harmonic trap
$V_t=M\sum_{\eta=x,y,z}\nu_\eta^2\eta^2/2$
($\nu_\eta=\omega\lambda_\eta$),
the equations for variational parameters $q_\eta$
are equivalent to the classical motion of a particle
(with coordinate $q_\eta$) inside an effective potential
\begin{eqnarray}
&&U(q_x,q_y,q_z)\nonumber\\
&=&\sum_{\eta=x,y,z}\left({\hbar^2\over 2M q^2_\eta}
+{M\omega^2\over 2}\lambda^2_\eta q^2_\eta\right)
+{Ng_0\over (2\pi)^{3/2}q_xq_yq_z}\nonumber\\
&+&N{g_2\over (2\pi)^{3/2}}{1\over q_xq_yq_z}\int d{\vec r}
\,{1-3{\cos}^2\theta\over r^3}\,e^{-\sum_\eta{\eta^2/2q^2_\eta}},
\label{eqv}
\end{eqnarray}
where $N$ is the number of condensate atoms.
The equilibrium location ($q_\eta^0$) of Eq. (\ref{eqv})
then determines condensate size, while linearized shape
oscillation frequencies are determined by the second order
derivative $U_{\eta\eta'}(q_x,q_y,q_z)$ evaluated at $q_\eta^{0}$,
For a cylindrically symmetric trap ($\lambda_x=\lambda_y=1$),
we take $q_x=q_y=q_r$ and $\lambda_z=\lambda$.
The last integral in (\ref{eqv}) as well as
$U_{\eta\eta'}$ all become analytically computable \cite{yi2}.
The resulting $U_{\eta\eta'}$ matrix is symmetric
($U_{\eta\eta'}=U_{\eta'\eta'}$), diagonalization of which gives
the three mode frequencies $\nu_a=\sqrt{U_{11}-U_{12}}$ and
\begin{eqnarray}
\nu_{b,c} ={1\over
\sqrt 2}\Big [&&U_{11}+U_{12}+U_{33}\nonumber\\
&&\pm\sqrt{(U^2_{11}+U^2_{12}-U^2_{33})^2+8U^2_{13}} \Big]^{1/2}.
\end{eqnarray}
\begin{figure}
\includegraphics[width=2.75in]{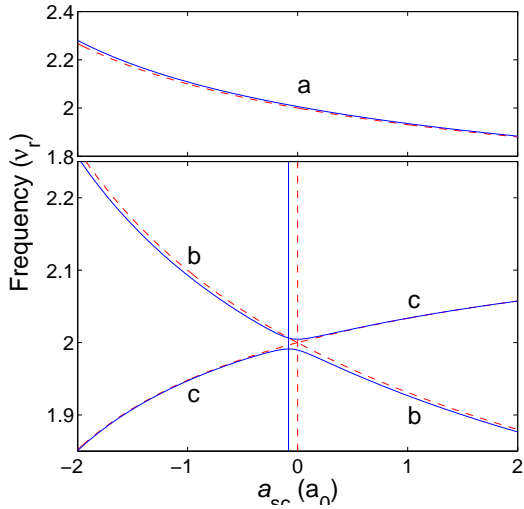}
\caption{Mode frequencies for $N=10^4$ and $\lambda=1$. Solid
(dashed) lines are results excluding (including) dipole
interactions. $a_0$ stands for Bohr radius.} \label{fig3}
\end{figure}

In the JILA experiments \cite{e2,e3} with $^{85}$Rb in the
$|F=2,M_F=2\rangle$ state, the valance electron spin gives rise to
an aligned magnetic dipole moment of $\mu=2\mu_B/3$ ($\mu_B$ being
the Bohr magneton). We take the radial frequency to be
$\nu_r=17.35$ (Hz), in addition to a tunable $a_{\rm sc}$
and trap aspect ratio $\lambda$ as in the experiment \cite{e3}.
Our results confirm that spin dipole effects become detectable
in terms of shifts of the shape oscillation frequencies.
We first report variational calculated results as the
analytic formulae obtained allow for a careful analysis of the
underline physics.
\begin{figure}
\includegraphics[width=2.75in]{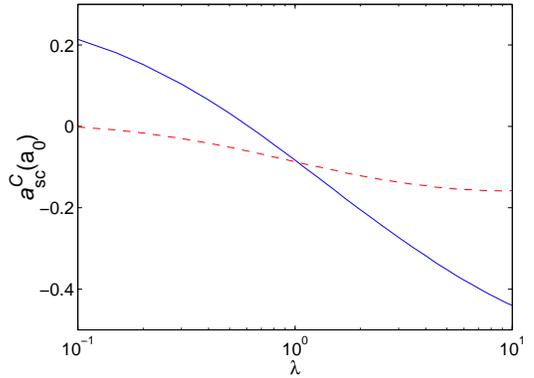}
\caption{The $\lambda$ dependence of the critical
scattering length $a_{\rm sc}^C$ for $N=10^4$.
The overall trend agrees with the variational result for
$a_{\rm eff}$ (in dashed line) derived earlier in Ref. [8].} \label{fig4}
\end{figure}

Figure \ref{fig3} shows the three mode frequencies for
$N=10^4$ in a spherical trap ($\lambda=1$). The solid and
dashed lines denote respectively mode frequencies
without and with the dipolar interaction.
The vertical lines show the critical values of $a_{\rm sc}^C$
when the mode character switching
$b\leftrightarrow c$ occurs. This mode
switching occurs whenever the overall mean-field
condensate interaction changes from
repulsive to attractive or {\it vice versa}.
When squeezed along the radial direction,
mode c (b) became predominantly excited for an overall
attractive (repulsive) condensate as atoms are
pulled in (pushed out) along the z-direction.
Without the spin dipole interaction, this mode switch
always occurs at $a_{\rm sc}=0$. The anisotropic
dipole interaction affects the overall condensate
mean field and the stability depending on the trap aspect ratio
$\lambda$ \cite{yi1,santos}. Depending on
the configuration of the dipoles, dipole-dipole interactions
can be either attractive or repulsive. For two dipoles,
if they were placed in a plane perpendicular to their
polarization ($\uparrow\uparrow$), they repel each other.
On the other hand, they attract each other if
they were placed along the direction of their polarization ($\rightarrow\rightarrow$).
For a cloud of trapped dipoles one therefore expects that repulsive
interaction increases as one increases $\lambda$, which leads to
increased condensate stability.
In Fig. \ref{fig4}, we display the $\lambda$ dependence
of the critical scattering length $a_{\rm sc}^C$ as obtained
from the variational calculation.
We find that the effective scattering
length $a_{\rm eff}$ (introduced earlier by us in \cite{yi1})
provides a reasonable approximation to the sign of
the overall condensate mean field. In terms of
the actual values, however, the results in the figure show that
$a_{\rm eff}$ differs significantly from $a_{\rm sc}^C$
when $\lambda$ deviates from the neighborhood of unity.

\begin{figure}
\includegraphics[width=2.75in]{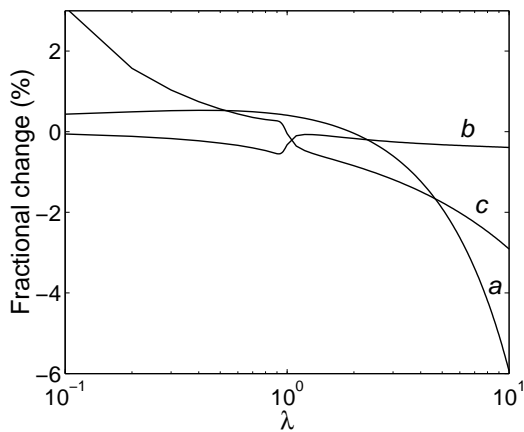}
\caption{Fractional change (\%) of mode frequencies for $a_{\rm
sc}=-1\,(a_0)$ and $N=10^4$. Similar results are obtained for
$a_{\rm sc}=0$ and $1\,(a_0)$.} \label{fig5}
\end{figure}

\begin{figure}
\includegraphics[width=2.75in]{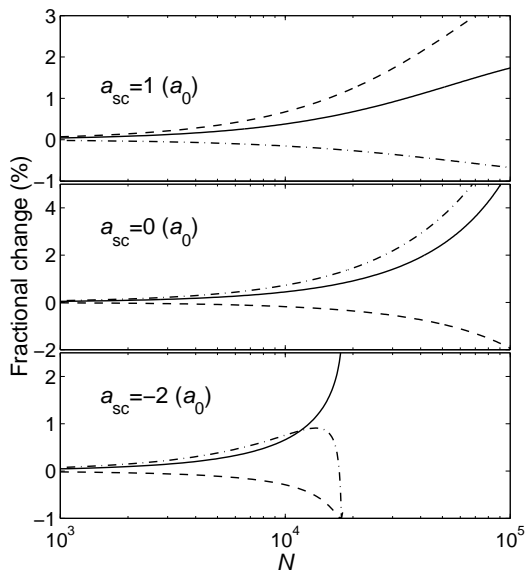}
\caption{Atom number dependence of the dipole interaction induced
shape oscillation frequency shifts.
Mode a, b, and c are labelled respective with solid, dashed,
and dot-dashed lines.} \label{fig6}
\end{figure}

In Fig. \ref{fig5}, we show the $\lambda$-dependence of
the dipole induced fractional changes to the mode frequencies
for a condensate with $10^4$ atoms at $a_{\rm sc}=-1\,(a_0)$.
For strongly prolate or oblate traps, the shifts
are in a few percentage range even with such small
numbers of atom.
Figure \ref{fig6} summarizes our results for
the atom number dependence of the dipole induced frequency
shifts for the trap aspect ratio $\lambda=6.8/17.35$ \cite{e3}.
It was shown earlier \cite{yi2},
at increased values of $N$, a dipolar condensate always collapses
irrespective of the sign of $a_{\rm sc}$. The Gaussian ansatz
(\ref{gausan}) becomes questionable near collapse as reflected
in the seemly divergent results when N is increased.

These results (Figs. \ref{fig3}-\ref{fig6})
clearly show that dipolar effects are detectable
in the current $^{85}$Rb BEC setup, given the extraordinary
capability of $0.1\%$ frequency measurement \cite{pri}.
To confirm the validity of the variational calculations,
we have invested considerable effort in
an exact numerical method based on the time dependent
Gross-Pitaevskii equation \cite{keith2}.
By applying external periodic forcing terms
described by the potential
$$
V_F(\vec r)=\sum_{\eta=x,y,z} V_\eta\left\{
e^{-{[\eta+\Delta_\eta(t)]^2\over 2w^2}}+
e^{-{[\eta-\Delta_\eta(t)]^2\over 2w^2}}\right\},
$$
%(see Fig. \ref{fig7})
with
$\Delta_\eta(t)=\eta_0+\delta_\eta\sin(\Omega_\eta t+\phi_\eta)$
to the condensate ground state, selected shape oscillations
become predominantly excited, as first numerically
implemented by Ruprecht {\it et al.} \cite{keith2}.
%\begin{figure}
%\includegraphics[width=2.in]{fig7.eps}
%\caption{Deformed (bare) trap potentials.} \label{fig7}
%\end{figure}

\begin{figure}
\includegraphics[width=2.75in,height=3.5in]{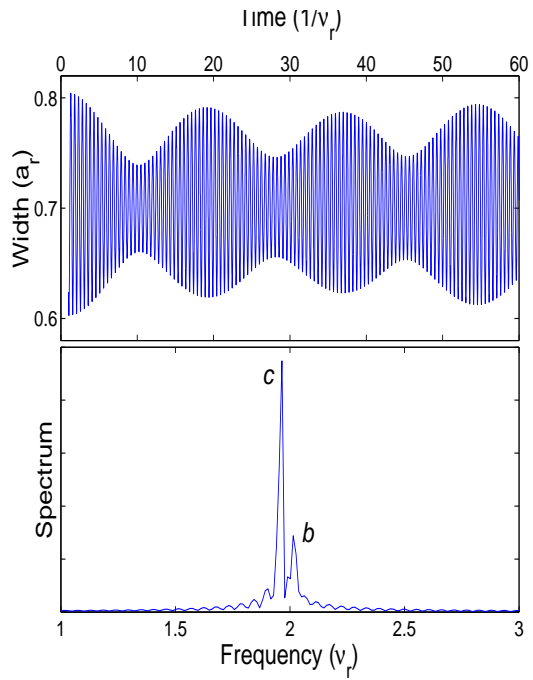}
\caption{Typical numerical results for condensate width and its
corresponding Fouier transformed signal. Mode $b$ and $c$ are
clearly identified. The parameters are $a_{\rm sc}=0$,
$\lambda=1$, and $N=40000$.} \label{fig8}
\end{figure}

After a selected duration $T$, typically several periods of
the trap radial oscillation,
the forcing term $V_F$ is tuned off. The free
propagation of the time dependent Gross-Pitaevskii equation is
continued, and the dynamic condensate width $\sqrt{\langle
\eta^2(t)\rangle}$ is sampled. The shape oscillation frequencies
are then identified by taking the Fourier transformation of
$\sqrt{\langle \eta^2(t)\rangle}$. A typical result from this
calculation is given in Fig. \ref{fig8}, which shows
remarkably clear signal. By varying $V_\eta$, $\eta_0$,
$\delta_\eta$, and $\Omega_\eta$, our results are
self-consistently checked, i.e. to be independent of all
parameters involved as they should be in the small amplitude
oscillation limit and to be numerically accurate.
For the experimental parameter of $\nu_r=17.35$ (Hz) and
$\nu_z=6.8$ (Hz). We have computed the shape oscillation
frequencies numerically for $a_{\rm sc}=0,1,-2$ ($a_0$), and
compared them with the variational results in Table
\ref{t1}. The quoted errors for the numerical results
shown in the tables mainly come
from the limited sampling window in time. Further improvement
is hard as the calculations are time consuming and
numerical accuracies become difficult to control at long times.
\begin{table}
\caption{Mode frequencies (in unit of $\nu_r$) with dipolar interaction}
\begin{tabular}{cccc}
$a_{\rm sc}$($a_0$) &mode &numerical &variational\\ \hline
$-2$ & $\begin{array}{c}b\\c\end{array}$
&$\begin{array}{c}2.0010\pm 0.0040\\0.8578\pm 0.0085\end{array}$
&$\begin{array}{c}2.0016\\0.8657\end{array}$ \\ \hline
$0$ &$\begin{array}{c}b\\c\end{array}$ &$\begin{array}{c}1.9959\pm
0.0028\\0.7874\pm 0.0022\end{array}$
&$\begin{array}{c}1.9964\\0.7895\end{array}$\\ \hline
$+1$ &$\begin{array}{c}b\\c\end{array}$ &$\begin{array}{c}0.7705\pm
0.0060\\1.9974\pm 0.0032\end{array}$
&$\begin{array}{c}0.7675\\1.9974\end{array}$ \\ \hline
\end{tabular}
\label{t1}
\end{table}
We find that the variational results are
consistent with the exact numerical results. In order to affirm
such mode frequency shifts are indeed from the dipolar interaction,
rather than a mis-calibration of $a_{\rm sc}$, we also need to
validate the variational approach in the absence of the
dipole interaction. This was explored earlier in Ref. \cite{zoller},
where the numerical and variational results of the condensate
dynamics were compared. Our results are presented
in Table \ref{t2}. It convincingly proves that
dipolar efforts reported in the Table \ref{t1}
is due to the physics of spin dipole interaction.
\begin{table}
\caption{Mode frequencies (in unit of $\nu_r$) without dipolar interaction}
\begin{tabular}{cccc}
$a_{\rm sc}$($a_0$)&mode&numerical&variational \\ \hline $-2$
&$\begin{array}{c}b\\c\end{array}$ &$\begin{array}{c}2.0073\pm
0.0025\\0.8546\pm 0.0051\end{array}$
&$\begin{array}{c}2.0075\\0.8591\end{array}$\\\hline $+1$
&$\begin{array}{c}b\\c\end{array}$ &$\begin{array}{c}0.7664\pm
0.0054\\2.0004\pm 0.0024\end{array}$
&$\begin{array}{c}0.7622\\2.0004\end{array}$\\ \hline
\end{tabular}
\label{t2}
\end{table}

In conclusion, we have studied low energy shape oscillations
of a trapped dipolar condensate. Using a magnetic field dependent
Feshbach resonance to tune the s-wave scattering length to around
zero, we have shown that the weak spin dipole interaction
becomes detectable as shifts to shape oscillation frequencies
under currently available experimental conditions \cite{e2,e3}.
These shifts grow with the number of trapped atoms, and
are around $1\%$ level with as few as $10^4$ atoms.
We have also independently verified the accuracy of the
variational calculation by developing a rigorous
numerical approach for the three predominant shape oscillation modes.
Near a Feshbach resonance, significant atom loss might occur
as found in the Na experiment \cite{mit2}. The subsequent damping
could lead to a broadening of the shape oscillation thus masking
the direct observation of the proposed spin dipole effects.
Fortunately for $^{85}$Rb atoms, the condition of $a_{\rm sc}(B)=0$
corresponds to far off resonance on the high B-field side, where
impressive controls have been demonstrated without any significant
loss \cite{e3}.
Over the last few years, mean field theory has proven to be
remarkably successful when applied to BEC physics.
The s-wave $g_0$ contact pseudo-potential has made the
concept of scattering length $a_{\rm sc}$ widely popular.
It is often said that scattering length is the only relevant
atomic parameter since the net interaction effect scales
as $Na_{\rm sc}/a_{\rm ho}$ for a harmonically trapped atomic
condensate, where $a_{\rm ho}$ is the trap size.
Dipole interactions as discussed in this article
points to interesting physics beyond the s-wave.
Within the current experimental regime of
dilute atomic gases, mean field theory remains applicable
and allows for the calculation of dipolar induced
shifts of collective excitation frequencies.
Successful experimental verification of our predictions
will shed new light on atomic BEC.

We thank Drs. C. E. Wieman and E. Cornell for helpful discussions.
This work is supported by the NSF grant No. PHY-9722410.

\end{document}